\documentclass[twocolumn,english,aps,prl,twocolum,groupedaddress,bibnotes,amsmath,amssymb,floatfix]{revtex4-1}
\usepackage[utf8]{inputenc}
\usepackage[english]{babel}
\usepackage{amsmath,amsfonts,amssymb}
\usepackage[T1]{fontenc}
\usepackage{url}
\usepackage[colorlinks=black,citecolor=blue,linkcolor=black]{hyperref}

\usepackage{amsmath}
\usepackage{amsfonts}
\usepackage{amssymb}
\usepackage[version=3]{mhchem}

\usepackage{epstopdf}
\usepackage{graphicx}
\graphicspath{{./Figures/}}

\newcommand{\fref}[1]{Fig.~\ref{#1}}

\begin{document}

\title{Cherenkov radiation induced symmetry breaking in counter propagating \\dissipative Kerr solitons}

\author{Romain Bouchand}
\thanks{These two authors contributed equally}
\author{Wenle Weng}
\thanks{These two authors contributed equally}
\author{Erwan Lucas}

\author{Tobias J. Kippenberg}
\email[]{tobias.kippenberg@epfl.ch}
\affiliation{Institute of Physics, École Polytechnique Fédérale de Lausanne (EPFL), CH-1015 Lausanne, Switzerland}

\begin{abstract}
The process of soliton Cherenkov radiation (also known as dispersive wave emission) in microresonator frequency combs plays a critical role in generating broadband and coherent microcomb spectra. Here, we report the observation of symmetry breaking in the group velocity of counter-propagating dissipative Kerr solitons, induced by polychromatic soliton Cherenkov radiation. Results show that in the presence of higher-order dispersion, the temporal arrangement of a multi-soliton state affects its group velocity, an effect which originates from the interference between multiple radiative tails emitted by individual solitons. Experimentally, we bidirectionally pump a microresonator with laser fields of equal power and frequency, and observe lifting of the degeneracy between the repetition rates of the counter-propagating solitons. The observation of symmetry breaking despite symmetric pumping conditions not only shines new light on the impact of dispersive waves on dissipative Kerr soliton dynamics, but also introduces a novel approach to develop coherent dual-comb spectrometers based on microcombs.
\end{abstract}

\maketitle


\def \DWRep {\Delta \omega_{\rm rep}}
\def \DFRep {\Delta f_{\rm rep}}

\textit{Introduction}---Dispersive waves are of critical importance in the study of wave propagation in a wide range of fields including hydraulics \cite{whitham1965general}, solid mechanics \cite{sachse1978determination}, magnetic systems \cite{rogers2001role} and ecology \cite{frantzen2000spread}. In the optical domain, dispersive wave formation (also known as Soliton Cherenkov radiation)~\cite{akhmediev1995cherenkov} belongs to one of the most important nonlinear optical processes, and has played a decisive role in the development of self-referenced optical frequency combs~\cite{udem2002optical,hansch2006nobel} by enabling coherent octave spanning spectra to be attained via supercontinuum generation~\cite{ranka2000visible,dudley2006supercontinuum,skryabin2010colloquium,markos2017hybrid}. These dispersive waves are emitted when the soliton phase matches a spectrally separated wave, and lead to oscillatory tails at either the trailing or leading edge of the soliton.
In a similar vein, the observation of soliton Cherenkov radiation~\cite{brasch2017self,jang2014observation} in microresonator-based optical frequency combs (microcombs)~\cite{del2007optical,kippenberg2011microresonator}, has provided a method to generate broadband and coherent microcomb spectra via the formation of dissipative Kerr solitons (DKS)~\cite{kippenberg2018dissipative}. In this context, Cherenkov radiation has been utilized for achieving self-referencing~\cite{brasch2017self}, octave spanning dual dispersive waves~\cite{Spencer2018}, self-locking of soliton group velocity (i.e. soliton repetition rate) \cite{skryabin2017self}, and deterministic single-soliton generation \cite{bao2017spatial}. In addition, various works have shown that Cherenkov radiation may be used as a means of tuning the soliton group velocity and consequently the repetition rate of microcombs \cite{lucas2017detuning,yi2017single,cherenkov2017dissipative}, which on the one hand may facilitate the agile control and the full stabilization of microcombs, but on the other hand constitutes a noise transduction mechanism for low noise microwave generation~\cite{yi2017single,liu2019nanophotonic}.
 
Here we observe and explain how soliton Cherenkov radiations can lead to the symmetry breaking of the repetition rates of counterpropagating dissipative Kerr solitons in an optical microresonator. 
In contrast to earlier work that deliberately imposed an asymmetry in either power~\cite{joshi2018counter}, or laser frequency~\cite{yang2017counter} to lift the repetition rate degeneracy, we observe that, counter intuitively, even for \emph{degenerate} pump field parameters (i.e. power, polarization and frequency), the degeneracy in soliton group velocity can be lifted.
To explain this phenomenon, we focus our attention on the interference between multiple dispersive waves that are emitted by the individual solitons forming a multi-soliton bound states. 
Our analysis reveals that the inter-soliton separation variation results in varied dispersive wave intensities through polychromatic Cherenkov radiation interferences. As a consequence the soliton recoil~\cite{akhmediev1995cherenkov,milian2014soliton,matsko2016optical} and the soliton group velocity differ, causing symmetry breaking of the repetition rates in microcombs with a different optical dissipative structure. We observe qualitative agreement between the experimentally observed symmetry breaking and our numerical simulations.
The observation of symmetry breaking in the repetition rate, beyond highlighting novel nonlinear dynamics in DKS, constitutes a new method to generate frequency comb spectra for dual comb spectroscopy using a single \emph{degenerate} pump field, and within a single whispering gallery mode family.
\begin{figure*}[t!]
\centering
\includegraphics[width=1.95\columnwidth]{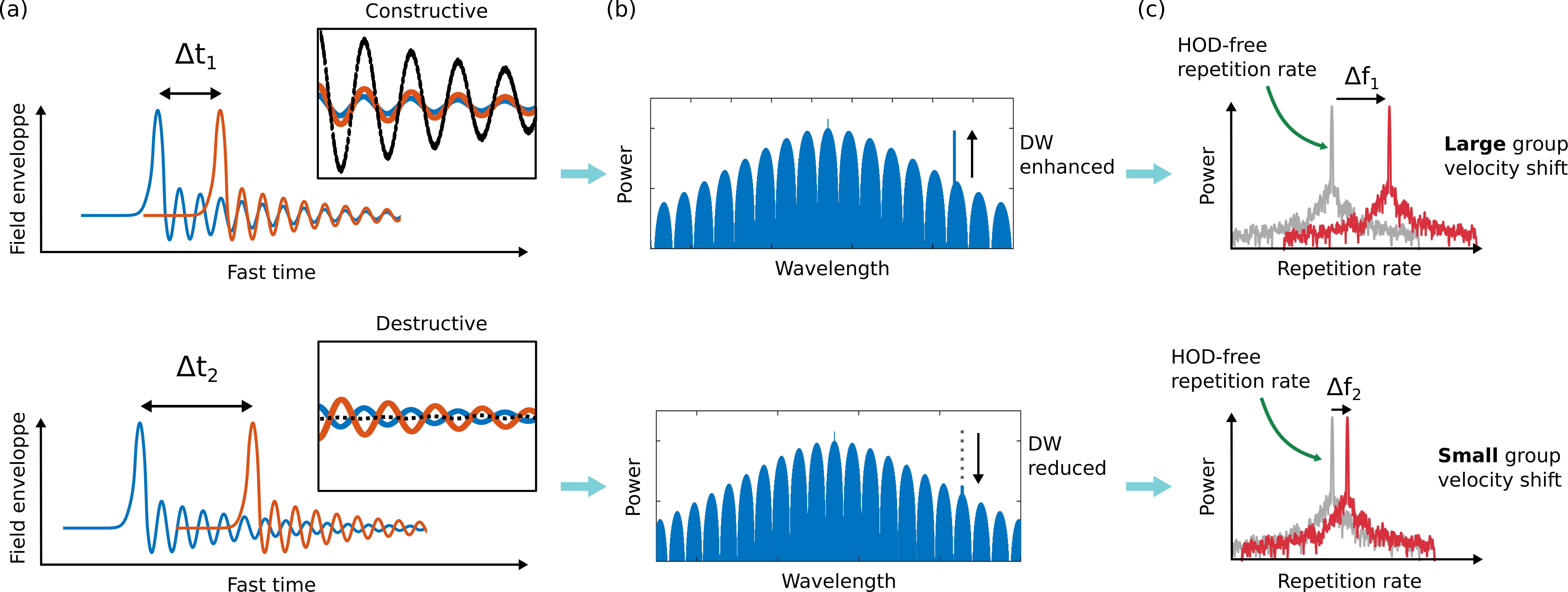} 
\caption{\textbf{Influence of Cherenkov radiation interference on the group velocity of a multi-soliton state.} (a) In the presence of higher-order dispersion (HOD), the solitons composing a two-soliton state each emit polychromatic Cherenkov radiation which interfere differently depending on their respective phase delay (i.e. the inter-soliton separation). (b) The interference between each soliton's Cherenkov radiation results in a selective enhancement or suppression of the given dispersive wave (DW) in the spectral domain. (c) Depending on the strength of the dispersive wave, the soliton group velocity is shifted by a variable amount from the HOD-free repetition rate (recoil effect).
As a result, when two-counter-propagating multi-soliton states driven by the same pump feature different temporal arrangement, the symmetry of the system is broken, and the soliton states travel at different group velocities. We note that the DW in (b) may seem misleading as, at first glance, it seems to convey that there is only a single-mode dispersive wave in the spectrum. However we emphasize that the effect we describe implies that there are always multiple dispersive features making the CR polychromatic and imposing aperiodic binding sites for the multi-soliton state, even if they are subtle and not easily visible in the spectrum. Nevertheless, the main contributor to soliton recoil is the strongest dispersive wave we depict.
}\label{fig:concept}
\end{figure*}

\textit{Influence of Cherenkov Radiation (CR) on the soliton group velocity}---A well-known phenomenon of DKS in optical resonators with higher-order dispersion is that resonant CRs are emitted when phase-matched to the dissipative soliton~\cite{coen2013modeling,brasch2016photonic}. For an optical microresonator with resonance frequencies:
\begin{equation}
\omega_\mu = \omega_0 + \sum_{j=1}\frac{D_j\,\mu^j}{j!}\quad,
\end{equation}
where $\mu$ is the mode index number and $D_j$ is the j-th order dispersion, the approximate condition is met at a mode of index $\mu_\mathrm{\footnotesize DW}$ when the integrated dispersion, $D_\mathrm{int}$, obeys $D_\mathrm{int}(\mu_\mathrm{DW})\buildrel \text{d{}ef}\over =\omega_{\mu_\mathrm{DW}} - (\omega_0 +D_1\,\mu_\mathrm{DW})=0$.
As a consequence, the dispersive waves can originate from higher order dispersion terms (i.e. $D_{i\,>3}$) of the mode family which supports the DKS~\cite{brasch2016photonic}, or can occur due to local mode crossings between different spatial mode families~\cite{yang2016spatial}, which can drastically modify the local dispersion profile.
Typically the CR manifests as a decaying modulation of the intracavity field background that is bound to the soliton~\cite{akhmediev1995cherenkov} and which provides a strong trapping potential for multi-soliton states~\cite{taheri2017optical}. As a consequence, for a multi-soliton state, different stable binding sites (and therefore inter-soliton separations) exist, leading both to interference patterns in the spectral envelope~\cite{brasch2016photonic,herr2014temporal} and to crystallized states~\cite{cole2017soliton}.
Moreover, the CR constitutes a loss mechanism where energy is either radiated from the trail or tail of the solitons and induces a positive or negative shift of the soliton group velocity in a deterministic way (the ``recoil'' effect)~\cite{milian2014soliton,bao2017spatial, yi2017single}. 
In prior works, CRs have been typically treated as single-period oscillations on the intracavity CW background which always constructively interfere when multiple solitons coexist~\cite{bao2017spatial,parra2017interaction,vladimirov2018effect} thereby creating a \textit{periodic} potential trap for the solitons with equidistant binding sites. In this picture, every binding site for the multi-soliton state is equivalent and will lead to constructive interference of the CR spectral component, producing a given group velocity shift. However, CRs are fundamentally \textit{polychromatic} objects which contain multiple spectral components that interfere with each other and which jointly create \textit{aperiodic} potential traps~\cite{wang2017universal} with irregularly spaced binding sites. In turn, the interferences, which are not always constructive due to the aperiodicity of the potential, result in the selective enhancement or suppression of specific dispersive waves and the subsequent recoils. Alltogether, these recoils lead to a global shift in the group velocity which is function of the inter-soliton distance. The concept is shown in \fref{fig:concept} on the example of a two-soliton state, where two different temporal arrangements (i.e. separation) lead to different group velocity depending on the CR interference. We note that a similar aperiodic trapping effect was recently reported in the case of 2D soliton bound states~\cite{milian2018clusters}.
\textit{Model}--- The numerical model describing the intracavity pulse dynamics is the Lugiato-Lefever equation~(LLE)~\cite{lugiato1987spatial}:
\begin{multline}
\label{eq1}
\frac{\partial{A}}{\partial t} + i \sum_{j=2} \frac{D_j}{j !} (\frac{\partial}{i\partial \phi})^j A - i g {|A|^2 A} =\\ 
\left( { - \frac{\kappa}{2} + i(\omega_0 - \omega_p) } \right){A} + {\sqrt{\kappa _{\rm ex}} \cdot s_{\rm in}}
\end{multline}
where ${{A}}$ is the envelope of the intracavity field, 
$\phi$ is the angular coordinate in the co-rotating frame, $g$ is the single photon Kerr induced nonlinear frequency shift,
${\kappa}$ is the cavity decay rate, ${\kappa_{\rm ex}}$ is the external fiber coupling rate, and ${|s_{\rm in}|^2} = \frac{P_{\rm in}}{\hbar \omega_0}$ is the driving photon flux, where $P_{\rm in}$ is the power of the main pump. 
In the absence of higher-order dispersion, the solution to the LLE is well approximated by the superposition of stationary solitons, $A_{\rm sol}$, and a CW background, $A_{\rm CW}$~\cite{herr2014temporal}. However, when CR (originating from both higher-order dispersion and mode crossings) are taken into account, additional terms (which correspond to dispersive waves) are included in the solution. This can be expressed as~\cite{skryabin2017self}:
\begin{equation}
\label{eq2}
A(\phi) = A_{\rm sol} + A_{\rm CW} + A_{\rm CR}
\end{equation}
Here $A_{\rm CR} = \sum A_\mu\exp(i (|\mu|\phi - \theta_\mu)) $, $\mu$ is the resonance mode number of the positive (higher frequency relative to the pump) and negative Cherenkov radiation peaks (for the pumped resonance $\mu = 0$), and $\theta_\mu$ is the relative phase of each radiation mode.
When multiple CR waves are present, they interfere with each other, and depending on the phase distribution, generate a particular modulation of the intracavity field. This function, which is complicated to derive analytically due to the complex dispersion profile, imposes aperiodic binding sites for the solitons in the multi-soliton states and, in turn, modifies the phase distribution until a bound state is formed when such mutual interaction reaches equilibrium. As depicted schematically in \fref{fig:concept}, for each of these possible temporal arrangements, the CRs interfere differently, causing different spectral recoils in the two directions, and breaking the perfect symmetry of the system.

\begin{figure}[t!]
\centering
\includegraphics[width=0.99\columnwidth]{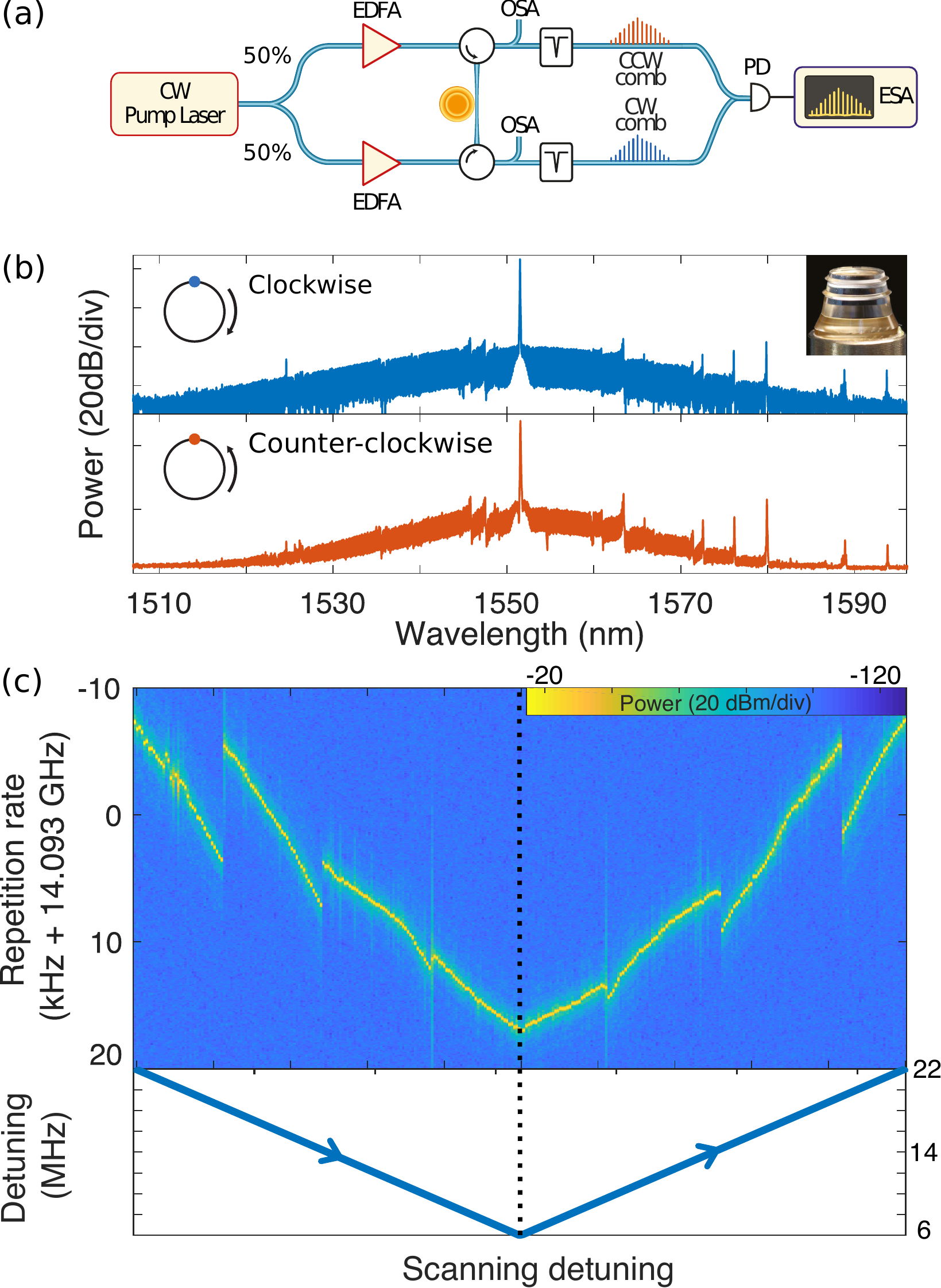} 
\caption{\textbf{Generation of degenerate counter-propagating soliton states in a single crystalline microresonator} (a) Experimental setup. Two single soliton states are excited by injecting counter-propagating pumps derived from the same laser in a tapered fiber via optical circulators. (b) Optical spectra of the single soliton state obtained in the CW (blue) and CCW (orange) directions. (c) The light obtained from the combination of two counter-rotating soliton states is mixed in a coupler, detected on a photodiode, and analyzed with an electrical spectrum analyzer. We plot the evolution of the repetition rate while scanning the pump to cavity resonance detuning and observe only one common signal which indicates that the two repetition rates are always degenerate. This signal exhibits a few abrupt changes in frequency that are due to the bistability induced by the dispersive waves.
}\label{fig:0}
\end{figure}
\textit{Experiments}--We choose to demonstrate the symmetry breaking effect in a configuration of two counter-propagating solitons in a single mode due to the perfect symmetry condition it offers. We generate DKS states in a crystalline MgF$_2$ microcavity~\cite{hofer2010cavity} ($n_0$=1.377, $n_2\sim$1.10$^{-20}$ m$^2$/W) that has been used in other works~\cite{lucas2017detuning,guo2017universal}. The microresonator has a free spectral range (FSR) of $D_1/2\pi$=14.09~GHz, the linewidth of the resonance that is pumped (at 1550~nm) is approximately $\kappa/2\pi\sim$ 100~kHz. The dispersion near 1550~nm is anomalous with $D_2/2\pi=$ 2~kHz and $D_3/2\pi=\mathcal{O}(\text{1~Hz})$. 
As shown in~\fref{fig:0}a, we pump a single spatial mode with clockwise (CW) and counter-clockwise (CCW) fields derived from the same laser source. By down-sweeping the laser frequency over the mode resonance we are able to generate soliton states in both directions, which can be analysed by observing the comb spectra and detecting the repetition rates as we tune the pump-resonance detuning. When both directions are single-soliton states, the repetition rates of both microcombs remain degenerate as the detuning is tuned over 16\,MHz, despite several jumps in repetition rates, which are attributed to the bistability induced by single-mode dispersive waves~\cite{yi2017single,weng2019spectral} (cf.~\fref{fig:0}). This is not surprising, as it has been reported that the back-scattering in microresonators can induce coupling between counter-propagating solitons and thus lead to a repetition rate locking effect~\cite{yang2017counter}. Typically, to lift the degeneracy, one needs to introduce a differential repetition rate shift to overcome the locking effect, which can be done by pumping the two directions with different intensities or frequencies \textit{via} the Kerr or Raman effects~\cite{yang2017counter,joshi2018counter}.
In an auxiliary experiment, we verified that this locking effect was present when we introduce differential repetition rate shifts through CR-induced recoil effects by pumping the two directions with distinct frequencies (see SI). In the following experiments however, the power and frequency of the two driving fields are kept equal to avoid any parasitic influence of the pumps asymmetry in the symmetry breaking mechanism.

\begin{figure}
\centering
\includegraphics[width=0.95\columnwidth]{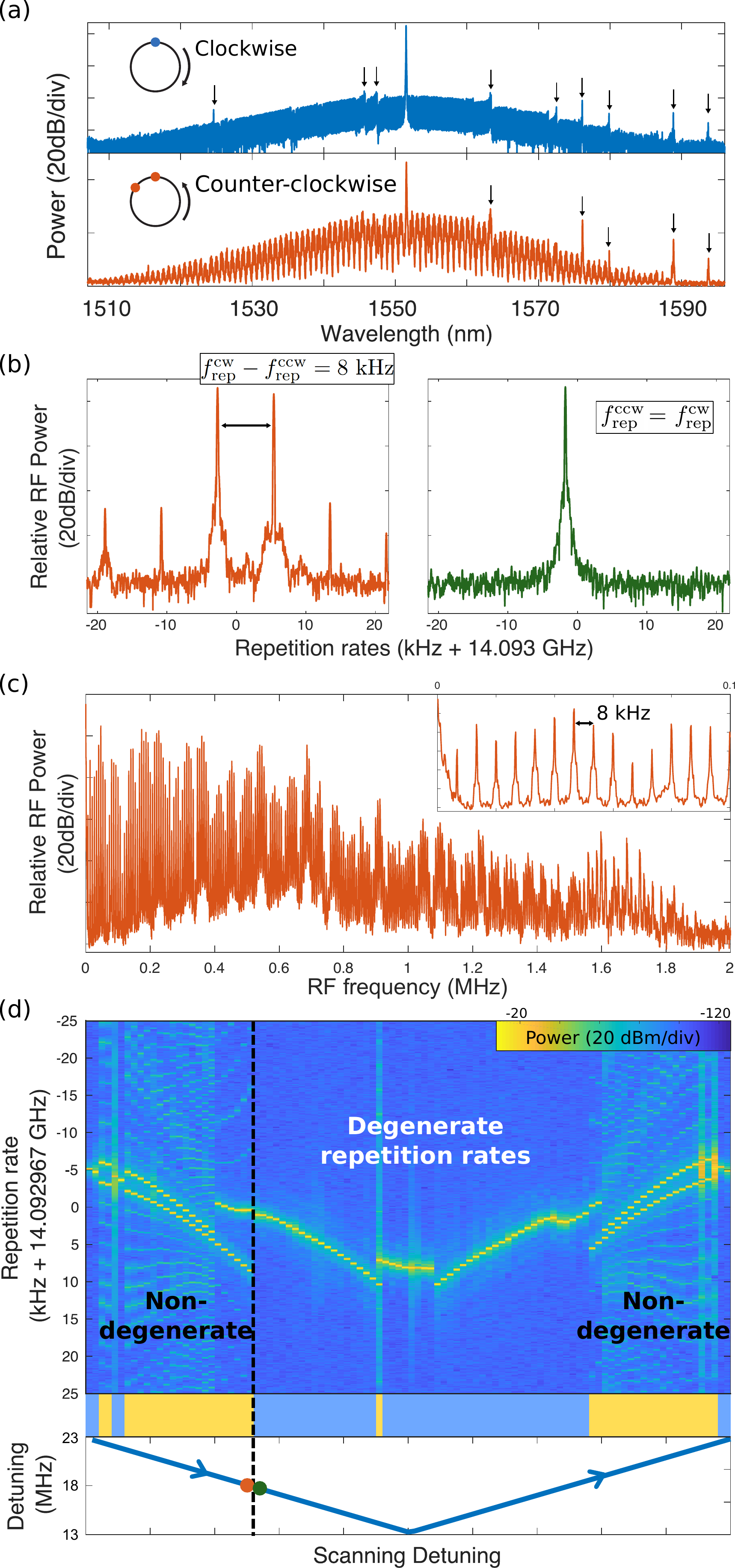} 
\caption{\textbf{Group velocity symmetry breaking between counter-propagating solitons.} (a) Spectra of a single soliton state in the CW direction (blue) and a double-soliton state in the CCW direction (orange). The most intense DWs are indicated by black arrows. (b) Detected repetition rate beatnote of the combined CW and CCW soliton pulse streams, obtained from photodetection of the mixed light from the two counter-propagating solitons (RBW is 500~Hz). The repetition rates are non-degenerate (left panel) or degenerate (right panel) due to the asymmetric interference of the polychromatic CR in the counter-propagating directions. (c) RF comb that we detect when the repetition rate are non-degenerate. Here, the line spacing is determined by the difference between the repetition rates of the two counter-rotating solitons. As the two pumps are degenerate, the RF comb starts from DC. (d) Evolution of the repetition rate degeneracy when scanning the pump detuning. The soliton group velocities are non-degenerate in the yellow regions.
}\label{fig:3}
\end{figure}

As a second experiment, a single-soliton state is generated in the clockwise direction and a double-soliton state in the counter-clockwise direction (see~\fref{fig:3}a) by state-switching to the desired DKS number~\cite{guo2017universal}. We then scan the detuning within the accessible soliton existence range, from 23~MHz to 13~MHz. As shown in \fref{fig:3}d, now that one of the two counter-rotating solitons is in a multi-soliton state, we are able to observe particular detuning regions for which the two solitons exhibit non-degenerate group velocities.
This alternation between degenerate and non-degenerate repetition rates depends on the differential strength of the dispersive wave-induced spectral recoils in the CW and CCW direction. The recoils are due to the joint effects of higher-order dispersion (mainly $D_3$) and multiple spatial mode-crossings. A thorough evaluation of the total effective repetition rate splitting between the two counter-rotating directions is then non-trivial to infer experimentally as it would require precise knowledge of all dispersive wave effects in both spectra. A simplified study is therefore conducted by considering only the differential recoil induced by the strongest dispersive wave at a particular detuning~(see~SI), which yields a reasonable quantitative estimation of the repetition rate splitting.

By choosing an appropriate detuning (e.g. 18~MHz as in Fig.~3a,3b (left panel) and 3c), we can make a dual-comb system from a single microresonator by generating counter propagating solitons with monochromatic pumping and equal power. Two distinct repetition rates are then detected on the ESA, see \fref{fig:3}b (left panel), separated by 8~kHz. The weak peaks on the sides (50~dB below the main signals) are modulation sidebands due to the solitons regularly colliding in the microresonator. The dual-comb configuration is corroborated by the baseband structure observed on the ESA (see \fref{fig:3}c). It consists of a comb of equidistant radio-frequencies starting from DC with a line spacing imposed by the difference between soliton group velocities. The comb structure starts at DC and is an experimental evidence that the two Kerr frequency combs that are generated are pumped by the same laser field. We emphasize that this result is fundamentally distinct from previous attempts in the literature that were either using different whispering gallery mode families~\cite{lucas2018spatial}, or counter-propagating modes with asymmetric pumps (unequal powers~\cite{joshi2018counter} or unequal frequencies~\cite{yang2017counter}).
\begin{figure*}[t!]
\centering
\includegraphics[width=1\textwidth]{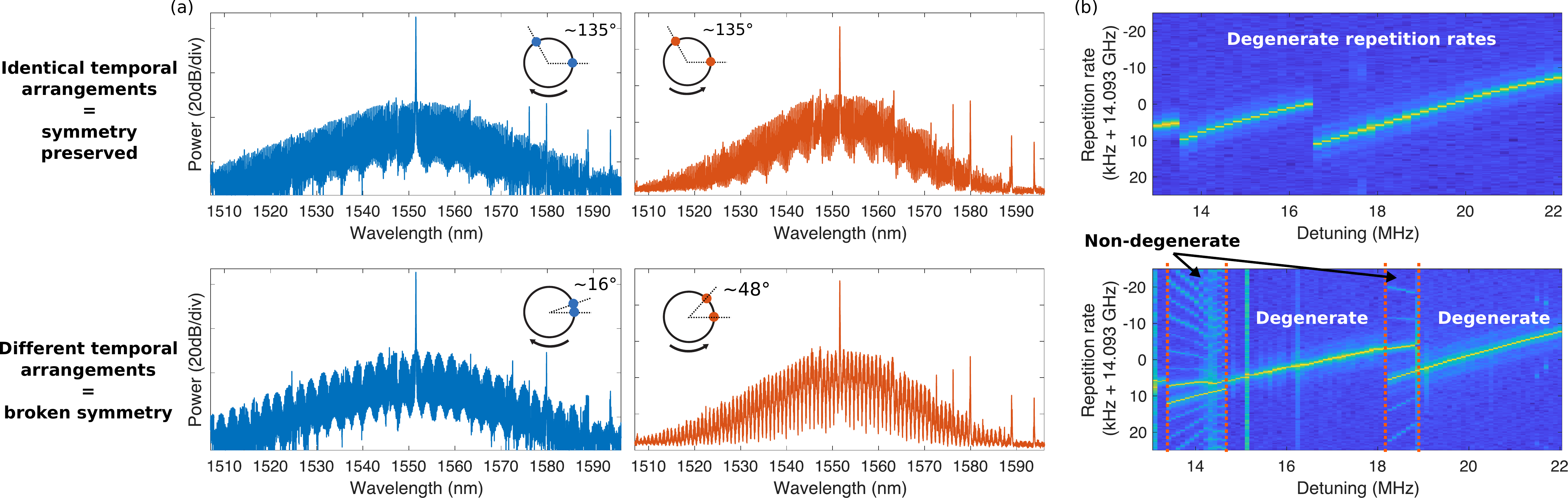} 
\caption{\textbf{Influence of multi-soliton state temporal arrangement on the symmetry of group velocities.} The upper panels correspond to counter-propagating two-soliton states with a nearly identical temporal configuration, while the lower panels correpond to the case of non-identical temporal separations. (a) Optical spectra of the two-soliton states in the CW (blue) and CCW (orange) direction. Insets: estimated temporal separation of the solitons in the two-soliton states. (b) Evolution of the repetition rate degeneracy when scanning pump to cavity resonance detuning.
}\label{fig:5}
\end{figure*}
This setup is then an ideal candidate for developing a dual-comb spectrometer where two combs with different repetition rates are used to probe the spectral information of an optical sample and map it to the RF domain~\cite{keilmann2004time,coddington2016dual,suh2016microresonator}. Here our repetition rate difference is in the 10~kHz range, and adjustable down to $\sim$1~kHz by acting on the pump-to-cavity detuning.
It's important to keep in mind that the RF comb which results from multi-heterodyne interference of the mixed combs has a repetition rate which is equal to the difference in individual comb repetition rates ($f_\text{rep}^{RF} = \Delta f_\text{rep} = f_\text{rep}^\text{cw} - f_\text{rep}^\text{ccw}$). This fact, combined with the exceptionally small $\Delta f_{rep}$ which we can achieve (along with the fact that pump degeneracy leads to the RF comb always beginning at DC), strongly relaxes the bandwidth requirements of the photodetector used, and leads to a very large optical-to-RF mapping factor of approximately $10^6$.

\textit{Simulations and discussion}--We corroborate the soliton group velocity symmetry breaking mechanism with extensive numerical simulations based on the LLE as discussed above. The parameters in the simulation are derived from the experimental setup presented above (see SI for details). The results confirm that, when higher-order dispersion effects are included, a multi-soliton state will adopt a particular (and seemingly random) temporal arrangement that results in a particular dispersive wave intensity which yields a particular soliton group velocity (through spectral recoil). However, if this hypothesis is valid, then any two counter-rotating multi-soliton states with the \emph{exact} same temporal arrangement must always have degenerate group velocities, due to the preserved symmetry. We experimentally verify this hypothesis by investigating two double-soliton states which exhibit (within our experimental precision) the same temporal distance. Here we scan the detuning in the usual manner, and show that in this case [\fref{fig:5}b (upper panels)], the repetition rate degeneracy is never lifted, irrespective of detuning~\footnote{Note that this case is in principle extremely unlikely given the formidable number of possible temporal arrangements. However, over a hundred of instances, we observed this case a couple of times, directly emerging from the chaotic modulation instability state of the intracavity field (see SI).}. Conversely, when the temporal separations are distinct (lower panels) we have a symmetry breaking of the soliton group velocities.
This confirms our previous hypothesis, and demonstrates that polychromatic CR interference is the underlying mechanism of the observed group velocity symmetry breaking.

In conclusion, we have observed and explained a novel mechanism that leads to symmetry breaking of counterpropagating soliton group velocities, despite identical pump frequency and power, due to polychromatic CR radiations. This counter-intuitive work sheds new light on the role of dispersive waves in multi-soliton state formation and, more practically, is an important step towards achieving compact monolithic spectrometers which require only one pump.\\

\begin{acknowledgments}
The authors thank Connor Skehan for his feedback on the manuscript. This publication was supported by Contract D18AC00032 (DRINQS) from the Defense Advanced Research Projects Agency (DARPA), Defense Sciences Office (DSO) and funding from the Swiss National Science Foundation under grant agreement No. 
~165933. W.~W. acknowledges support by funding from the European Union’s H2020 research and innovation programme under grant agreement No. 753749 (SOLISYNTH). E.~L. acknowledges the support of the European Space Technology Centre with ESA Contract No.~4000118777/16/NL/GM.
\end{acknowledgments}


\bibliography{TheReferences.bib}
\newpage
\onecolumngrid
\section{Supplementary material for: ``Cherenkov radiation induced symmetry breaking in counter propagating dissipative Kerr solitons''}
\twocolumngrid
\section{Experimental setup}
The light from a continuously tunable laser (Toptica CTL) is split in two paths, amplified in two indepedent EDFAs, frequency-shifted in two acousto-optic modulators (AOMs) and coupled evenescently to a MgF$_{2}$ microresonator in counter-propagating directions \textit{via} two fibered circulators connected at each end of a tapered fiber. The pump powers at the taper inputs are adjusted to the same power ($\sim 450$ mW). The third output ports of the circulators containing  the counter-propagating lights are partially sent to two optical spectrum analyzers (OSA) while the remaining parts are recombined in a coupler and sent to an electrical spectrum analyzer (ESA) and an oscilloscope. The countinuous wave (CW) laser is offset sideband locked from one mode of the microresonator using a standard Pound-Drever-Hall technique~\cite{drever1983laser,thorpe2008laser} in order to stabilize the effective detuning of the pump to the cavity resonance~\cite{lucas2017soliton,weng2019spectral}.
Dissipative Kerr soliton states are simultaneously generated in the two counter-propagating directions using the forward detuning method~\cite{herr2014temporal}. In order to stably access a detuning for which the two counter-propagating states are supported, we look simultaneously at the transmission from both rotating directions by detecting the mixed light on a photodetector, and adapt the detuning to land on the desired step after scanning the pump laser frequency.
\begin{figure}[b]
\centering
\includegraphics[width=0.95\columnwidth]{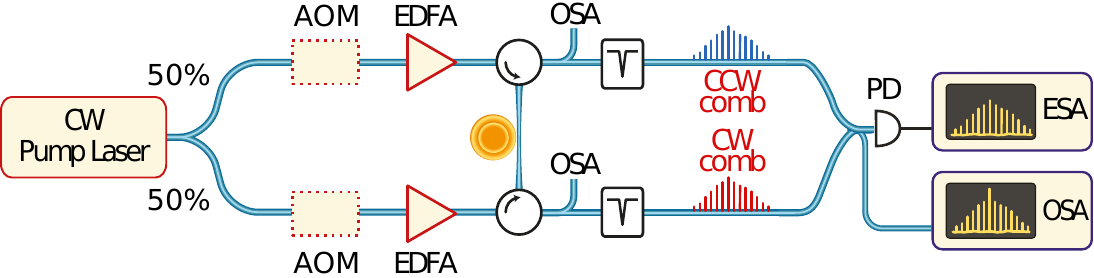} 
\caption{Experimental setup. AOM: Acousto-optic modulator; EDFA: Erbium-doped fiber amplifier; OSA: Optical spectrum analyzer; ESA: Electrical spectrum analyzer. CCW comb: counter-clockwise comb. CW comb: clockwise comb.}\label{fig:SI0}
\end{figure}
We performed two main experiments: in the first one the AOMs drive frequencies are adjusted in order to induce a small frequency shift between the two pumps (typically several kHz), while in the second one (described in the main text) the AOMs are not present and a single drive is used. The microresonator used in our experiments is a crystalline MgF$_2$ microcavity ($n_0$=1.377, $n_2\sim$~1.10$^{-20}$ m$^2$/W) that has been used in other works~\cite{lucas2017detuning,lucas2017breathing,guo2017universal,guo2017inter}. The microresonator has a free spectral range (FSR) of $D_1/2\pi$=14.09~GHz, the linewidth of the resonance that is pumped is approximately $\kappa/2\pi\sim$ 100~kHz. The dispersion near 1550~nm is anomalous with $D_2/2\pi=$ 2~kHz and $D_3/2\pi=\mathcal{O}(\text{1~Hz})$. The effective mode area is $A_\text{eff}\sim$150~$\mu$m$^2$.

\section{Non-degenerate pumps}
\begin{figure}[b!]
\centering
\includegraphics[width=0.85\columnwidth]{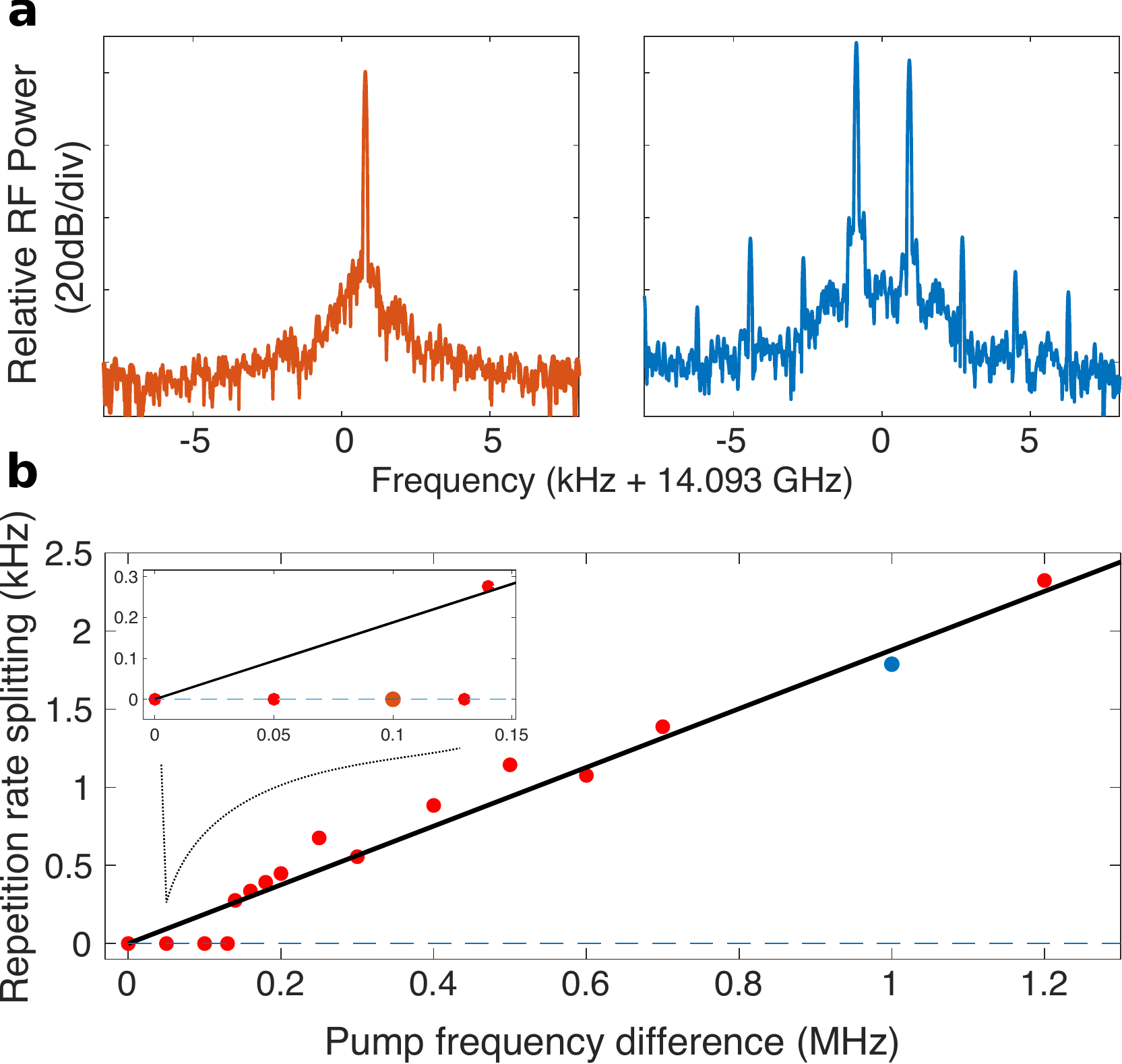} 
\caption{Group velocity symmetry breaking when pump frequencies are non-degenerate. (a) Repetition rate of the mixed CW and CCW DKS for two distinct pump frequency differences. When the pumps detuning is 100 kHz the two repetition rates are degenerate (blue) while when the pumps detuning is 1~MHz the degeneracy is lifted and two peaks corresponding to the two rotating solitons are visible (orange). (b) Representation of the difference between the CW and CCW repetition rates with respect to the pump frequency difference. The inset is a close-up of the plot for low pump frequency difference, it shows that the soliton repetition rates are locked when the pump frequency difference is below 140~kHz.}\label{fig:SI1}
\end{figure}
In this experiment, the two counter-propagating soliton states are generated using different pump frequencies ($\delta \omega_\text{\tiny Pump}=\omega_\text{\tiny CW}~-~\omega_\text{\tiny CCW}\neq~0$) hereby introducing an asymmetry in the system. The configuration is similar to what was reported by Yang et al. in the SiO$_2$ microresonator platform~\cite{yang2017counter}. The typical results obtained are displayed in~\fref{fig:SI1}. We observe that when the pumps are detuned from each other by more than 140~kHz, the repetition rates of the two counter-propagating solitons split and two repetition rate peaks are observable after detecting the mixed light from the two directions on the same photodiode. The lift of degeneracy is induced because of the different effective detunings of the pumps respective to the cold cavity resonance. However, here unlike previous works in SiO$_2$ platforms, the tunability of the repetition rates with the detuning is not due to the Raman self-frequency shift~\cite{karpov2016raman,yi2016theory} because the Raman gain in MgF$_2$ material is spectrally narrow (the strongest phonon mode close to room temperature is typically a lorentzian having a linewidth of 250~GHz~\cite{porto1967raman,grudinin2013impact}). In our case the tunability originates from the higher-order dispersion effects in the microresonator that leads to a modification of the group velocity of the soliton (and then the repetition rate) with the detuning~\footnotetext{A rapid estimation of the effect of the third order dispersion for a detuning of 1~MHz yields a repetition rate difference of $\delta f_r = \delta\omega D_3/(3\times 2\pi D_2)\sim$100~Hz, the factor ten difference between the experimental and theoretical values can be explained in part by the error on the measurement of $D_3$, by the effect of mode-crossings and by the contibution from the Kerr shock~\cite{bao2017soliton}.}\cite{yang2016spatial,cherenkov2017dissipative}.  When the pump frequency difference is smaller than 140~kHz, the repetition rate splitting becomes small enough so that it cannot counteract the trapping effect of the backscattered light and the two solitons will lock to each other~\cite{yang2017counter}, as is confirmed by the unique RF beat at 14.09~GHz in the RF spectrum (blue curve in the \fref{fig:SI1} (a)). This backscattering-induced soliton interlocking can be seen as a very efficient injection locking where the whole comb spectra participate to the locking. Besides, in this experiment we carefully choose the reference detuning $\delta\omega$ (the laser detuning when the two pumps are degenerate in frequency) so that no strong mode-crossing is disturbing the repetition rate splitting through abrupt changes in the repetition rate~\cite{matsko2016optical,yang2016spatial}. We note that we find experimentally a locking range close to what was observed in a SiO$_2$ platform~\cite{yang2017counter}. This can appear surprising as the scattering effects should be much less pronounced in MgF$_2$ materials, however we have a significantly higher quality factor (10$^9$ instead of 200$\times$10$^6$) that can compensate for this parameter and yield the same locking range.
The asymmetric cross-phase modulation of the CW and CCW pump light can also yield a lift of degeneracy of the two counter-propagating DKS repetition rates when the pump power ratio is not equal to unity. In this case the differential pumps powers causes differential non-linear phase shifts in the two directions, which subsequently results in a difference in the repetition rates~\cite{del2017symmetry}. This phenomenon has been used to generate dual-comb from a single Si$_3$N$_4$ microresonator with a single pump frequency and repetition rate splitting at the MHz level were obtained~\cite{joshi2018counter}. However in that case, the locking range due to backscattering was more than several MHz so that a power ratio was required to be typically superior to 10\% to obtain a repetition rate splitting.
In our case, given the material property of the MgF$_2$ microresonator compared to Si$_3$N$_4$ chips ($n_2$ 25 times weaker, FSR 14 times lower and the effective mode area 150 times larger) this effect is extremely minute and a two-fold difference between the CW and CCW pumps would only induce a repetition rate splitting of a few tens of Hz that would not counteract the solitons interlocking due to backscattering. In our main experiment with a single pump frequency, we nevertheless keep the power difference within $\pm$5\% to suppress the aforementioned effect.

\section{Degenerate pumps: simplified evaluation of the differential repetition rate shift}
\begin{figure}[b]
\centering
\includegraphics[width=0.85\columnwidth]{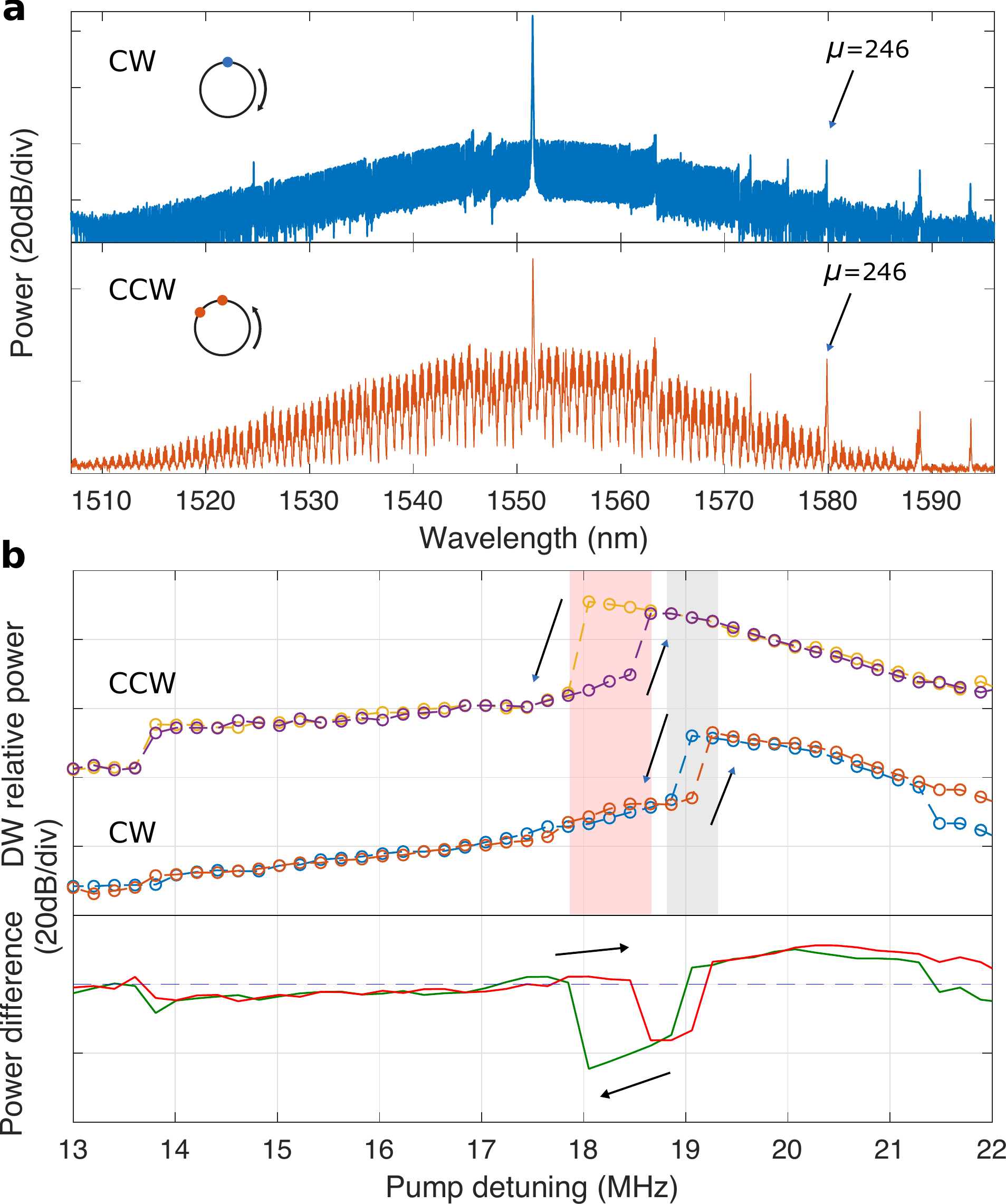} 
\caption{Spectra of the solitons generated in the CW (higher panel) and CCW (lower panel) direction when monochromatically driving the crystallline microresonator for a detuning of 19~MHz.  They correspond to a single soliton and a double soliton state, respectively (i.e. the same case depicted in the Fig.3 of the main text). They exhibit typical spatial mode-crossing-induced dispersive waves, among which we emphasize the mode -246 that present a different evolution in the two combs (b) Hysteretic evolution of the dispersive wave power for the mode -246 in the CW (top) and CCW (lower) direction, respectively. The arrows indicate the direction of the hysteresis cycles. The pink (grey) shaded area represents the detuning range for which the hysteretic behaviour appear for the CW (CCW) direction. The lower panel represents the variation of the difference in power between the dispersive wave (at mode number -246) between the two soliton combs when scanning the detuning up and down. The blue dotted line corresponds to an equal power of the dispersive wave between the two counter rotating solitons.
}\label{fig:SI2}
\end{figure}

As described in the main text, when we obtain a lift of degeneracy of the counter-rotating soliton group velocities due to the presence of a multi-soliton state with CR interference, a thorough evaluation of the repetition rates splitting would require taking into account the effect of every dispersive wave in each comb (see~\fref{fig:SI2}) and calculate the spectral recoils and the subsequent repetition rate shifts using a reformulation of the analytical expressions derived in \cite{yi2017single}:
\begin{equation}\label{eq:1}
\delta f_\text{rep}^\text{\tiny CP} = -\displaystyle\sum_{i \in \text{DW}} \mu_i\frac{D_2\,\kappa_i}{D_1\,\kappa}\;\left( \frac{P_i^\text{\tiny cw}}{E^\text{\tiny cw}}-\frac{P_i^\text{\tiny ccw}}{E^\text{\tiny ccw}} \right) 
\end{equation}
where $\mu_i$ is the number of the mode in which the dispersive wave is generated (counted from the pump), $\kappa$ is the linewidth of the pumped mode generating the soliton, $\kappa_i$ is the linewidth of the crossing mode which is inducing the dispersive wave, $E^\text{\tiny cw (ccw)}$ is the circulating soliton energy and $P^\text{\tiny cw (ccw)}_i$ is the power of a given dispersive wave for the CW (CCW) solitons.
The analysis can be singificantly simplified by noting that an intense single-mode dispersive wave is dominating in the comb spectra. The formula can then be evaluated for a single dispersive wave and yield an estimation of the repetition rate splitting. We look at the detuning region close to 18~MHz where we know from the experiment that the dispersive wave in the mode $\mu=$ -246 undergoes a sheer change in power in the CW comb but not in the CCW comb that is causing a repetition rate splitting of $\delta f_\text{rep}\sim$~8~kHz (see main text). Evaluating \eqref{eq:1} with the parameters of our microresonators and calculating $E$ and $P_{-246}$ from the experimental optical spectrum of the two combs, we find: $
\delta f_\text{rep} \sim \frac{\kappa_{-246}}{\kappa} \times 2.5\, [\text{kHz}]$\footnote{The absolute power of the soliton combs and of the dispersive waves need not to be known as only the power ratio intervenes in the equation.}.
The power loss rate of the crossing mode causing the dispersive wave is unknown but the value required to match the experimental data is: $\kappa_{-246}\,\sim 3.2\,\kappa$, which is totally acceptable given that the crossing mode linewidth is slightly broader than the intrinsic linewidth of the pumped mode (we deliberately choose to pump the mode family with the highest quality factor). The important result behind this is that the right order of magnitude is found to explain the repetition rate splitting between the counter-propagating solitons as induced by a differential spectral recoil due to the strongest dispersive wave.
\section{Generation of the soliton states}
In order to generate the different soliton states in the two directions we first sweep the diode laser frequency and use the forward tuning method~\cite{herr2014temporal}. Once the two counter-propagating solitons are generated, they are usually multi-soliton states. To obtain fewer soliton numbers in one of the direction, we scan the pump-to-cavity detuning by sweeping the offset frequency of the Pound-Drever-Hall phase lock loop~\cite{weng2019spectral}. The solitons number can then iteratively be decreased by using the backward tuning method~\cite{guo2017universal}.
When generating dual two-soliton states with this method they usually exhibit different temporal arrangements. However, in the striking case where we obtained totally symmetric two-solitons states (with the exact same temporal arrangement) it was always obtained directly with the forward tuning method, emerging from the chaotic modulation instability state. It can be related to some synchronization mechanism occuring during the chaotic state, which would represent an interesting feature to investigate in future works.

\section{Dispersive wave intensity versus inter-soliton separation: simulation and experiment}\label{sec:smdw}
\begin{figure*}[p]
  \centering{
  \includegraphics[width = 1.9\columnwidth]{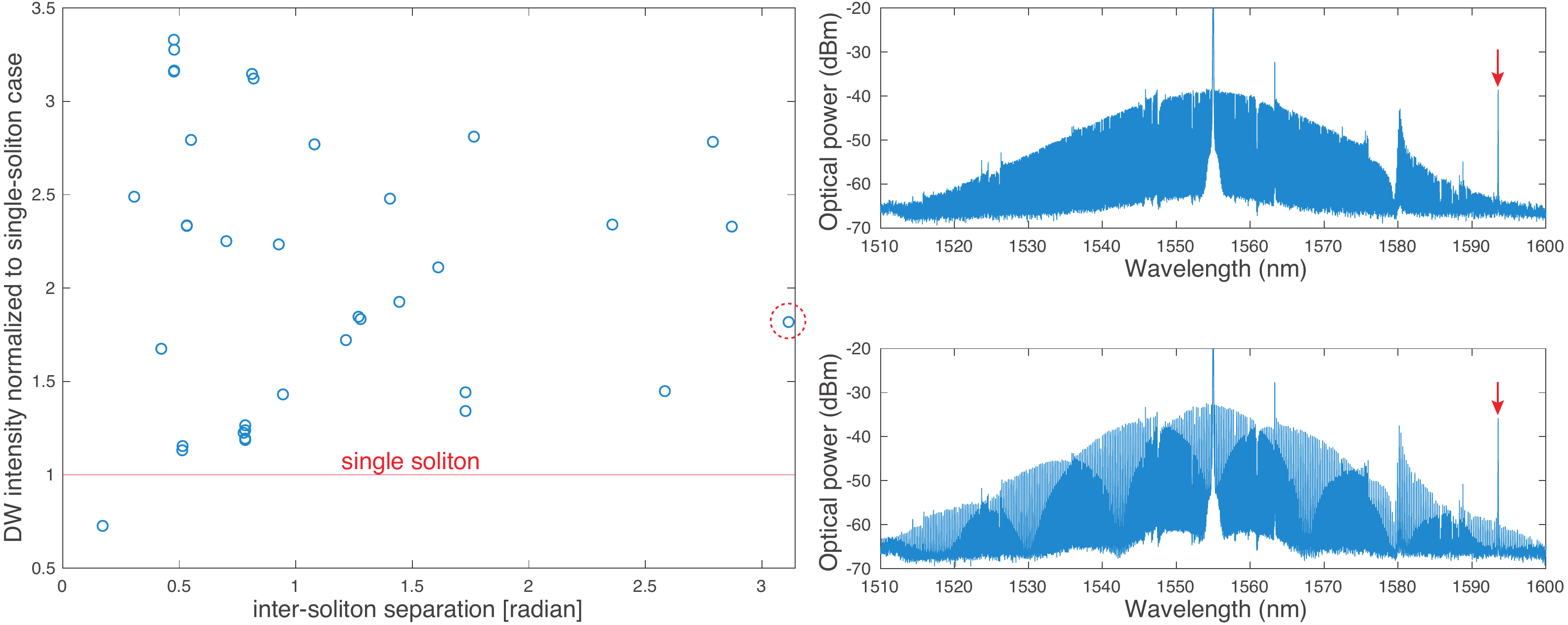}
  }
  \caption{\textbf{Experimentally measured single-mode dispersive wave intensity with respect to the inter-soliton separations for 37 randomly generated two-soliton states.} In the left all the SMDW intensities are normalized to the intensity for single-soliton state. A spectrum of a single-soliton state is shown in the upper-right, and a spectrum of a two-soliton state which corresponds to the data point in dashed red circle is shown in the lower-right. The SMDW at 1594\,nm is indicated by the red arrows.}
  \label{fig:SI3}
\end{figure*}

We experimentally generated two-soliton states in a repetitive manner and each time measured the intensity of a strong single-mode dispersive wave (SMDW) on the long-wavelength wing of the comb spectrum. The SMDW is indicated by a red arrow in the spectra presented in \fref{fig:SI2}. We also derived the inter-soliton separations between the two solitons from the interference patterns of the comb spectra. After having generated 37 such soliton states while keeping the pump power and the effective detuning constant  we plotted the measured SMDW intensity versus the inter-soliton separation in~\fref{fig:SI3}. The results show that the intensity of the SMDW is erratic with respect to the inter-soliton separation. We then measured the SMDW intensity for 6 different single-soliton states and we found that the measured intensity is constant in all 6 events, as indicated by the red line in~\fref{fig:SI3}. The comparison shows that in two-soliton states, the SMDW intensity is randomly distributed between the levels of total destructive interference (which corresponds to 0 in normalized intensity) and total constructive interference (which corresponds to 4 in normalized intensity).
\begin{figure*}[p]
  \centering{
  \includegraphics[width = 1.9\columnwidth]{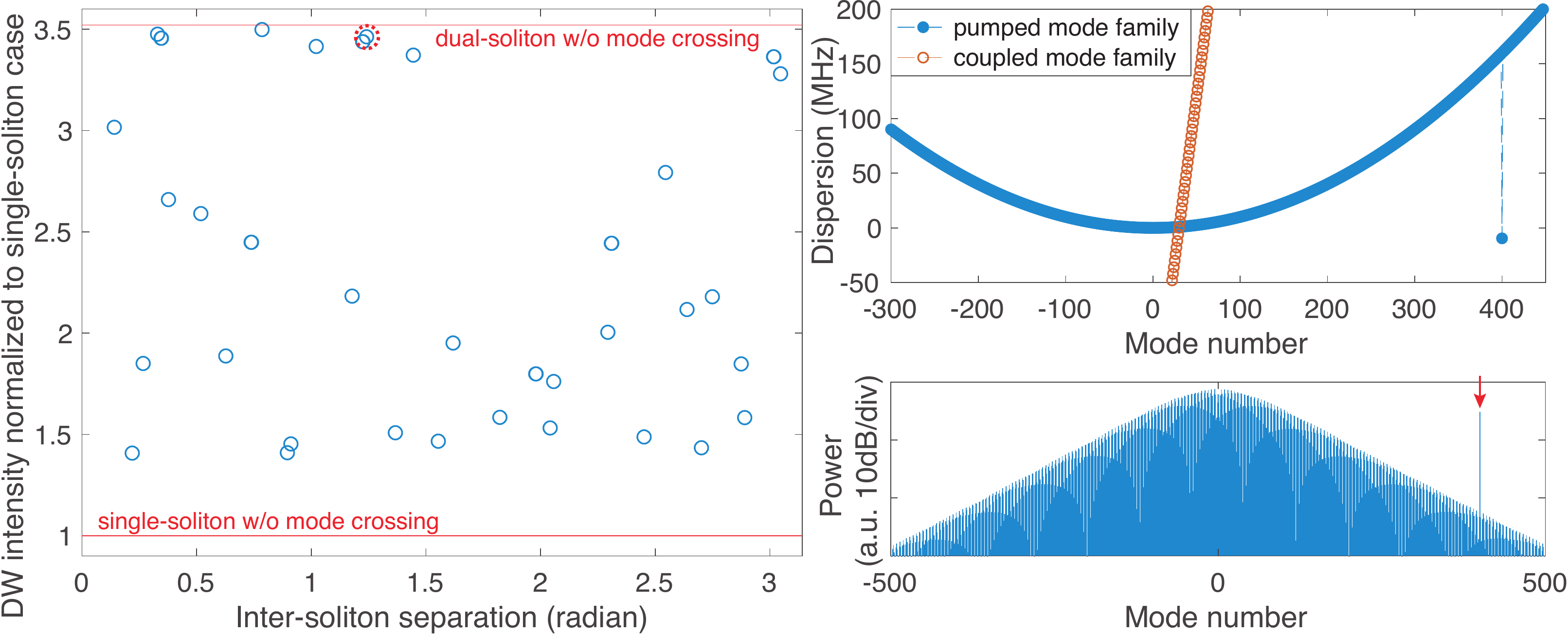}
  }
  \caption{\textbf{Simulated single-mode dispersive wave intensity with respect to the inter-soliton separations for 50 randomly seeded two-soliton states.} In the left all the SMDW intensities are normalized to the intensity for a single-soliton state (without mode crossing effect). In the upper-right the integrated dispersion of the pumped mode family is displayed in blue circles, while the relative resonance frequencies of the coupled mode family are presented in red circles. In the lower-right the comb spectrum of one of the randomly seeded two-soliton states which corresponds to the data point in the red circle in the left is presented, with the red arrow indicating the SMDW.}
  \label{fig:SI4}
\end{figure*}
\begin{figure*} [t!]
  \centering{
  \includegraphics[width = 1.9\columnwidth]{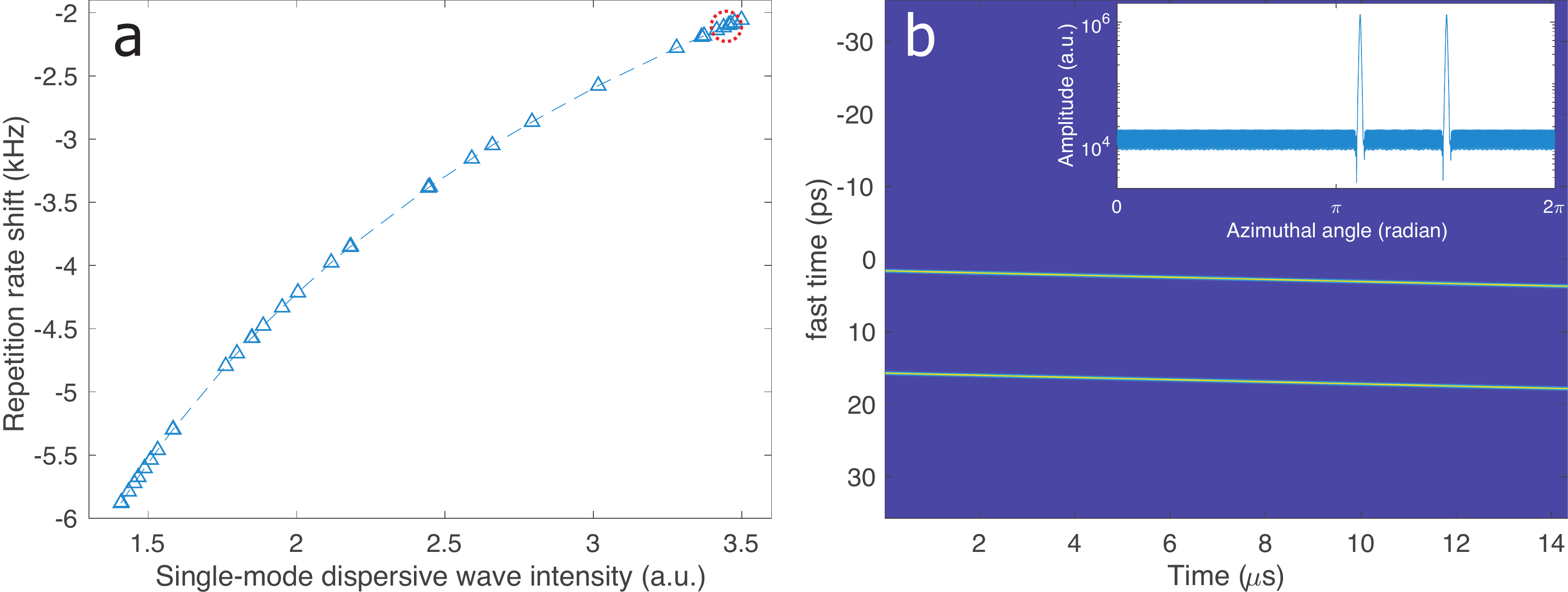}
  }
  \caption{\textbf{Relation between soliton repetition rate shift and single-mode dispersive wave intensity (from simulations).} (a) Simulated data from 50 randomly seeded two-soliton states. (b) Evolution of the micoresonator intracavity field in presence of one of the randomly seeded states (which corresponds to a data point in the red circle in (a)). The inset shows a shot of the intracavity field at a particular time.}
  \label{fig:SI5}
\end{figure*}
To verify our postulation that the randomly distributed SMDW intensities are caused by the interference of polychomatic dispersive waves, we carried out numerical simulations based on the Lugiato-Lefever equation (LLE). In the simulation the second-order dispersion coefficient is set as $\frac{D_2}{2\pi}=2$\,kHz. We introduce a SMDW by changing the resonance frequency of the corresponding mode (deviation of $\sim 48\kappa$ of the resonance of mode 400). We also use the approach detailed in other works~\cite{yi2017single,guo2017inter}, to introduce a local dispersion disruption due to the effect of mode coupling. The integrated dispersions of the pumped and coupled mode families are displayed in Fig.\,\ref{fig:SI4} (right panel). We set the coupling factor $\frac{g}{2\pi} = 600$\,kHz and the loss rate of the coupled mode family $\frac{\kappa_{\rm{c}}}{2\pi}=800$\,kHz. Then we initiate the split-step simulation of two-soliton states by seeding the solitons with random inter-soliton separation. After a period of at least ten photon-decay times when the group velocities of the solitons and the inter-soliton separations settle, the SMDW intensity is derived from the simulated optical spectrum. These observations show that different inter-soliton separations in a multi-soliton state lead to different SMDW intensities, that we attribute to interference between polychromatic dispersive waves emitted by individual solitons.

From the same set of simulated data, we also calculated the DKS repetition rate shift by numerically fitting the intracavity motion of solitons. Fig.\,\ref{fig:SI5}\,(a) shows the relation between the repetition rate shift and the intensity of the SMDW. We observe that the repetition rate shift increases monotonically as the SMDW intensity rises, which is due to the increased soliton frequency recoil with stronger dispersive wave. Fig.\,\ref{fig:SI5}\,(b) shows the evolution of the intracavity field for one randomly seeded state. The inset shows a shot of the intracavity field, in which we also observe the dispersive wave as field oscillations of the CW background. This simulation confirms that different SMDW intensities will induce different shifts in the group velocity of a multi-soliton state. This can cause a symmetry breaking between the group velocities of counter-propagating solitons sharing the exact same pump, provided that the soliton states are not identical in each direction (i.e. different number of solitons or different inter-soliton separations). 

\end{document}